\documentclass[12pt]{article} 
\usepackage{graphicx} 
\def \b{{\cal B}} 
 
\def \bea{\begin{eqnarray}} 
\def \beq{\begin{equation}}

\def \eea{\end{eqnarray}} 
\def \eeq{\end{equation}}

\def\lsim{\mathrel{\rlap{\lower3pt\hbox{$\sim$}}\raise2pt\hbox{$<$}}}
\def\gsim{\mathrel{\rlap{\lower3pt\hbox{$\sim$}}\raise2pt\hbox{$>$}}}

\begin{document} 
\begin{flushright} 
SLAC-PUB-13184 \\ 
EFI 08-04 \\ 
arXiv:0803.3229 [hep-ph] \\ 
March 2008 \\ 
\end{flushright} 
\renewcommand{\thesection}{\Roman{section}} 
\renewcommand{\thetable}{\Roman{table}}
\centerline{\bf EXAMINATION OF FLAVOR SU(3) IN $B,~B_s \to K \pi$ DECAYS
\footnote{To be submitted to Physics Letters B.}} 
\medskip
\centerline{Cheng-Wei Chiang}
\centerline{\it Department of Physics and Center for Mathematics and
  Theoretical Physics}
\centerline{\it National Central University, Chungli, Taiwan 320,
  R.O.C. and}
\centerline{\it Institute of Physics, Academia Sinica, Taipei, Taiwan
  115, R.O.C.}
\medskip 
\centerline{Michael Gronau} 
\centerline{\it Stanford Linear Accelerator Center, Stanford University,
Stanford, CA 94309 \footnote{On sabbatical leave from the Physics Department,
Technion, Haifa 32000, Israel.}}
\medskip 
\centerline{Jonathan L. Rosner} 
\centerline{\it Enrico Fermi Institute and Department of Physics,
  University of Chicago} 
\centerline{\it Chicago, IL 60637, U.S.A.} 
\bigskip 
\begin{quote} 
We study a relation between the weak phase $\gamma$ and the rates and CP
asymmetries of several $K \pi$ decays of $B^+$, $B^0$, and $B_s$, emphasizing
the impact of the latter measurements.  Current data indicate large SU(3)
breaking in the strong phases or failure of factorization (including its
application to penguin amplitudes) in $K \pi$ modes of $B^0$ and $B_s$.
SU(3) and factorization only remain approximately valid if the
branching ratio for $B_s \to K^- \pi^+$ exceeds its current value of
$(5.27 \pm 1.17) \times 10^{-6}$ by at least 42\%, or if a parameter $\xi$
describing ratios of form factors and decay constants is shifted from its
nominal value by more than twice its estimated error.
\end{quote}
 
\leftline{\qquad PACS codes:  12.15.Hh, 12.15.Ji, 13.25.Hw, 14.40.Nd} 
 
\bigskip 

Several methods have been proposed to measure the Cabibbo-Kobayashi-Maskawa
(CKM) phase $\gamma$ from $B$ meson decays into $D K$ final states
\cite{Gronau:1991dp,Atwood:1996ci,Giri:2003ty} and in charmless strange final
states using flavor SU(3) symmetry \cite{Neubert:1998pt,Gronau:2001cj,%
Chiang:2003rb,Sun:2003wn,Chiang:2005kz}.  Ref.~\cite{Gronau:2000md} proposed
using $B^+ \to K^0 \pi^+$, $B^0 \to K^+ \pi^-$, and $B_s \to K^- \pi^+$, the
last two related by the U-spin symmetry $d \leftrightarrow s$, to obtain
$\gamma$.  (A recent analysis employing this method is described in Ref.\
\cite{Fleischer:2007hj}.)  Ignoring ${\cal O}(\lambda^2)$ terms in the $B^\pm
\to K^0 \pi^\pm$ decay amplitude,\footnote{Ref.~\cite{Chiang:2000za}
illustrates the effect of a ${\cal O}(\lambda^2)$ term from the penguin
amplitude, but a color-suppressed penguin amplitude of the same order
is not included.} where $\lambda = 0.2257$ \cite{Wolfenstein:1983yz,PDG},
$\gamma$ is obtained from the ratios of decay widths.

The ratio of contributions of $B_s$ and $B^0$ to the $K^\pm \pi^\mp$ final
state in proton-antiproton collisions has recently been reported with improved
accuracy by the CDF Collaboration \cite{CDF}.  The result is $(f_s/f_d) \b(B_s
\to K^- \pi^+)/\b(B^0 \to K^+ \pi^-) = 0.071 \pm 0.010({\rm stat.})
\pm 0.007({\rm sys.})$, where $f_s/f_d$ is the ratio of production fractions of
$B_s$ and $B^0$.  Given the world averages \cite{HFAG} $f_s = (10.4 \pm
1.4)\%$, $f_d = (39.8 \pm 1.0)\%$, and $B(B^0 \to K^+ \pi^-) =
(19.4 \pm 0.6) \times 10^{-6}$, this implies $\b(B_s \to K^- \pi^+) =
(5.27 \pm 0.74 \pm 0.90) \times 10^{-6}$.  We include this result along
with direct CP asymmetries in $B^0 \to K^+ \, \pi^-$ and $B_s \to K^- \, \pi^+$
to solve for $\gamma$, the strong phases, and the ratio between tree and
penguin amplitudes.  We find in general a two-fold ambiguity in the solutions
for weak and strong phases.  Moreover, we find a large SU(3)-breaking effect
either between the strong phases or between the magnitudes of
strangeness-conserving and strangeness-changing amplitudes, given the present
experimental situation \cite{CDF}.

We review the method proposed in Ref.~\cite{Gronau:2000md}.  Employing U-spin
symmetry, the decay amplitudes of the relevant modes are
\begin{eqnarray}
\label{GRamp-a}
A(B^+ \to K^0 \, \pi^+) &=& P ~, \\
\label{GRamp-b}
A(B^0 \to K^+ \, \pi^-) &=&
  T \, e^{i(\delta_d+\gamma)} + P ~, \\
\label{GRamp-c}
\xi A(B_s \to K^- \, \pi^+) &=& 
  \frac1{{\tilde \lambda}} T \, e^{i(\delta_s+\gamma)}
     - {\tilde \lambda} P ~,
\end{eqnarray}
where the explicit $t$-quark dependence is removed using CKM unitarity.
Here $T$ and $P$ denote ``tree'' and ``penguin'' amplitudes, proportional to
the CKM factors $V_{us}^*V_{ub}$ and $V_{cs}^*V_{cb}$, respectively.  The
parameter ${\tilde \lambda} \equiv \left|V_{us}/V_{ud}\right| \simeq 0.2317$
using $\lambda = 0.2257$ \cite{PDG} and $V_{ud} = \sqrt{1 - \lambda^2}$.  We
also include an overall SU(3)-breaking factor
\begin{eqnarray} \label{eqn:xi}
\xi \equiv 
\frac{f_K F_{B^0 \pi}(m_K^2)}{f_{\pi} F_{B_sK}(m_{\pi}^2)}
\frac{m_{B^0}^2 - m_{\pi}^2}{m_{B_s}^2 - m_K^2}
\end{eqnarray}
according to the factorization assumption for the amplitudes.\footnote{This
includes the assumption that the penguin and tree amplitudes scale in the same
way.  The consequence of relaxing this assumption will be explored.}
Its value is $0.97^{+0.09}_{-0.11}$ \cite{PDG,Khodjamirian:2003xk},
corresponding to almost exact SU(3).\footnote{We have assumed a vector
dominance pole model to extrapolate the form factors from the $q^2 = 0$ point
computed in Ref.~\cite{Khodjamirian:2003xk}.}  This should be compared with
global fits done within flavor SU(3) \cite{Chiang:2004nm,Chiang:2006ih}, which
associated the breaking factor $f_K/f_\pi \simeq 1.2$ with tree-type amplitudes
only.  In that case, the predicted branching ratios of the $B_s \to K^- \pi^+$
and $K^+ K^-$ modes \cite{Chiang:2006ih} agreed with the later experimental
measurements.  The relative strong phases between $T$ and $P$ are denoted by
$\delta_d$ and $\delta_s$ for $B^0 \to K^+ \pi^-$ and $B_s \to K^- \pi^+$,
respectively.  For consistency, terms of ${\cal O}(\tilde\lambda^2)$
have been ignored in these amplitudes.  Since interactions directly
involving the spectator quark are expected to be dynamically
suppressed, we also ignore their contributions.

Consider the charge-averaged ratios \cite{Gronau:2000md}
\begin{eqnarray}
R_d \equiv
\frac{\Gamma(B^0 \to K^+ \pi^-) + \Gamma({\overline B}^0 \to K^- \pi^+)}
{\Gamma(B^+ \to K^0 \pi^+) + \Gamma(B^- \to {\overline K}^0 \pi^-)} ~, \\
R_s \equiv
\frac{\Gamma(B_s \to K^- \pi^+) + \Gamma({\overline B}_s \to K^+ \pi^-)}
{\Gamma(B^+ \to K^0 \pi^+) + \Gamma(B^- \to {\overline K}^0 \pi^-)} ~,
\end{eqnarray}
and the CP-violating rate pseudo-asymmetries:
\begin{eqnarray}
A_d \equiv
\frac{\Gamma({\overline B}^0 \to K^- \pi^+) - \Gamma(B^0 \to K^+ \pi^-)}
{\Gamma(B^- \to {\overline K}^0 \pi^-) + \Gamma(B^+ \to K^0 \pi^+)}
& = & R_d A_{CP}(B^0 \to K^+ \pi^-) ~, \\
A_s \equiv
\frac{\Gamma({\overline B}_s \to K^+ \pi^-) - \Gamma(B_s \to K^- \pi^+)}
{\Gamma(B^- \to {\overline K}^0 \pi^-) + \Gamma(B^+ \to K^0 \pi^+)}
& = & R_s A_{CP}(B_s \to K^- \pi^+) ~.
\end{eqnarray}
Defining the ratio $r \equiv T/P$, we derive
\begin{eqnarray}
\label{eq:rd}
&&
R_d = 1 + r^2 + 2 r \cos\gamma \cos\delta_d ~, \\
\label{eq:rs}
&&
\xi^2 R_s = 
{\tilde \lambda}^2 + \left( \frac{r}{\tilde \lambda} \right)^2
- 2 r \cos\gamma \cos\delta_s ~, \\
\label{eq:ACPBd}
&&
A_d = 2 r \sin\gamma \sin\delta_d ~, \\
\label{eq:ACPBs}
&&
\xi^2 A_s = -2 r \sin\gamma \sin\delta_s ~.
\end{eqnarray}
Here we have ignored very small phase space differences.  Eqs.~(\ref{eq:ACPBd})
and (\ref{eq:ACPBs}) imply a simple relation between the strong phases:
\begin{eqnarray}
\label{eq:sinedelta}
\frac{\sin\delta_d}{\sin\delta_s}
= - \frac{A_d}{\xi^2 A_s}
= - \frac{R_d A_{CP}(B^0 \to K^+ \, \pi^-)}
         {\xi^2 R_s A_{CP}(B_s \to K^- \pi^+)}~.
\end{eqnarray}
Numerically, this ratio is $0.96 \pm 0.54$ according to the
data in Table~\ref{tab:exp}.

\begin{table}
\begin{center}
\begin{tabular}{lrc}
\hline\hline
Observable & Exp.\ Value & Ref.\ \\
\hline
$\b(B^+ \to K^0 \pi^+)$ & $23.1 \pm 1.0$ & \cite{HFAG} \\
$\b(B^0 \to K^+ \pi^-)$ & $19.4 \pm 0.6$ &\cite{HFAG} \\
$A_{CP}(B^0 \to K^+ \pi^-)$ & $-0.097 \pm 0.012$ & \cite{HFAG} \\
$\b(B_s \to K^- \pi^+)$ & $5.27 \pm 1.17$ & \cite{CDF} \\
$A_{CP}(B_s \to K^- \pi^+)$ & $0.39 \pm 0.17$ & \cite{CDF} \\
\hline\hline
\end{tabular}
\end{center}
\caption{Experimental values of observables used in this analysis.
  Branching ratios are charge-averaged and in units of $10^{-6}$.
  To convert their ratios to those of rates we use the lifetime
  ratios \cite{HFAG} $\tau(B^+)/\tau(B^0) = 1.071 \pm 0.009$ and
  $\tau(B_s)/\tau(B^0) = 0.939 \pm 0.021$.
\label{tab:exp}}
\end{table}

First, we consider the SU(3) limit where $\delta_d = \delta_s \equiv \delta$.
In this case, $\gamma$ and $\delta$ always appear in the combinations
$\cos\gamma \cos\delta$ and $\sin\gamma \sin\delta$ in Eqs.~(\ref{eq:rd}),
(\ref{eq:rs}), (\ref{eq:ACPBd}) and (\ref{eq:ACPBs}).  This set of equations
has the discrete symmetries (i) $\gamma \leftrightarrow \delta$ and $r$
invariant; (ii) $\gamma \to \gamma + \pi$, $\delta \to \delta + \pi$, and $r$
invariant; (iii) $\gamma \to \gamma + \pi$, $r \to - r$, and $\delta$
invariant; and (iv) $\delta \to \pi - \delta$, $\gamma \to \pi - \gamma$, and
$r$ invariant.  The amplitude ratio $r$ is negative according to the
factorization assumption.  In the following analysis, we therefore consider
only solutions with $0 \le \gamma \le 90^\circ$ and $r < 0$.  This still leaves
the two-fold ambiguity (i) mentioned above.

Eqs.~(\ref{eq:rd}) and (\ref{eq:rs}) give the absolute value of the
ratio between the redefined tree and penguin amplitudes
\begin{eqnarray}
\label{eq:absr}
|r| = {\tilde\lambda} \sqrt{\frac{R_d + \xi^2 R_s}{1 +
    {\tilde\lambda}^2} - 1} ~.
\end{eqnarray}
Using the experimental inputs listed in Table~\ref{tab:exp}, we have $R_d =
0.899 \pm 0.048$, $R_s = 0.260 \pm 0.059$, $A_d = 0.087 \pm 0.012$, and $A_s =  
-0.101 \pm 0.050$.  Eq.~(\ref{eq:absr}) implies $|r| \simeq 0.068 \pm 0.034$
with the SU(3) breaking factor $\xi$ included.  If $\xi$ is set to (1,~1.2),
$|r|$ increases to $(0.073 \pm 0.026,~0.106 \pm 0.024)$.  The condition
$R_d < 1$ demands $r \cos\gamma \cos\delta < 0$ according to Eq.~(\ref{eq:rd}).

The $B^0 \to K^+ \pi^-$ and $B_s \to K^- \pi^+$ rate asymmetries satisfy the
relation
\begin{eqnarray}
\Gamma(B_s \to K^- \pi^+) - \Gamma({\overline B}_s \to K^+ \pi^-)
= - \frac{1}{\xi^2} 
\left[ \Gamma(B^0 \to K^+ \pi^-) 
- \Gamma({\overline B}_d \to K^- \pi^+) \right]
\end{eqnarray}
by U-spin symmetry.  We can thus use $A_{CP}(B^0 \to K^+ \pi^-)$
to predict $A_{CP}(B_s \to K^- \pi^+) \simeq 0.35 \pm 0.12$.
This is consistent with the measured value in Table~\ref{tab:exp}.

As $\b(B^+ \to K^0 \pi^+)$ and $\b(B^0 \to K^+ \pi^-)$ have been determined
to about $5\%$, their current central values are not likely to vary much in the
future.  In contrast, $\b$ and $A_{CP}$ of $B_s \to K^- \pi^+$ have only been
measured by the CDF Collaboration for the first time.  The quoted value of
$\b(B_s \to K^- \pi^+)$ \cite{CDF} depends on the fragmentation fractions
$f_s$ and $f_d$ \cite{HFAG} (see also Ref.\ \cite{Aaltonen:2008zd}),
whose ratio carries a 14\% error.  (The total systematic
error on $\b(B_s \to K^- \pi^+)$, including this contribution, is 17\%.)
In the following, we discuss the
dependence of solutions on the central value of $\b(B_s \to K^- \pi^+)$.
As $\delta_s$ has been fixed to be the same as $\delta_d$, we omit $A_{CP}(B_s
\to K^- \pi^+)$ from the fit and predict its value from the fit parameters.
Errors and other measurements are kept at their current values.

\begin{figure}[t!]
\centering{\includegraphics[width=0.7\textwidth]{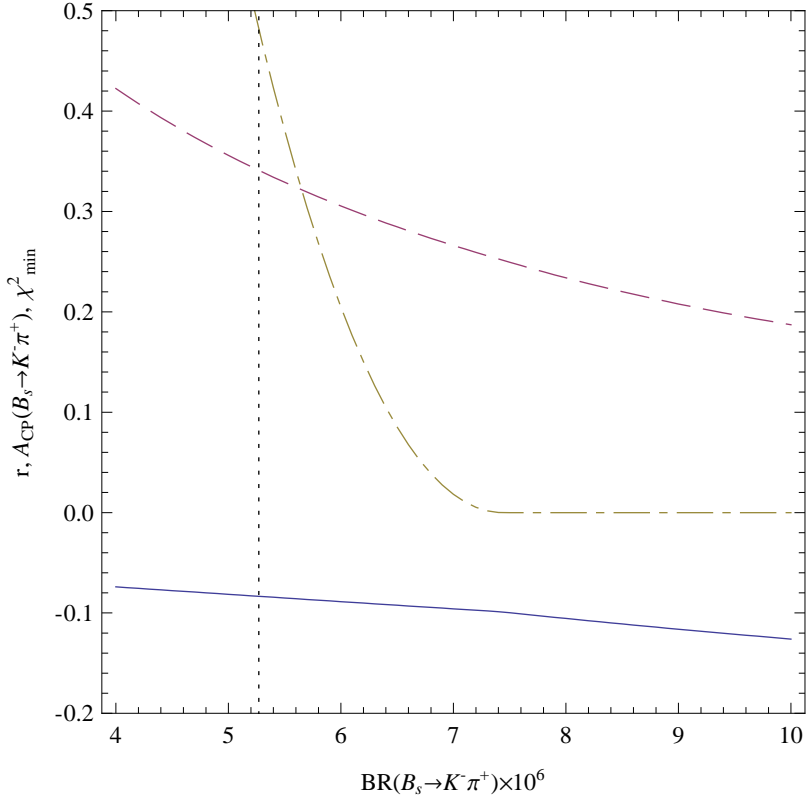}}
\caption{Behavior of solutions as a function of $\b(B_s \to K^-
  \pi^+)$, assuming $r<0$ and $\delta_d = \delta_s \equiv
  \delta$.  The solid, dashed, and dot-dashed curves represent $r$,
  preferred $A_{CP}(B_s \to K^- \pi^+)$, and $\chi^2_{\rm min}$,
  respectively.  The vertical dotted line indicates the current
  central value of $\b(B_s \to K^- \pi^+)$.}
\label{fig:rACP}
\end{figure}

Fig.~\ref{fig:rACP} shows the dependence of $r$ on $\b(B_s \to K^- \pi^+)$; the
$\chi^2$ for the fit to $R_d$, $R_s$, and $A_{CP}(B^0 \to K^+ \pi^-)$; and the
predicted $A_{CP}(B_s \to K^- \pi^+)$.  For $\b(B_s \to K^- \pi^+) \gsim 7.5
\times 10^{-6}$, a solution with strong phases satisfying exact SU(3) can be
obtained, as indicated by the vanishing $\chi^2_{\rm min}$ value.  This can be
attributed to an overall SU(3) breaking factor of at least $\xi \simeq 1.2$,
2.5 $\sigma$ from the central value $0.97$ given by factorization and echoing
the observation in Ref.~\cite{Chiang:2006ih} mentioned above.  For $\b(B_s
\to K^- \pi^+) < 7.5 \times 10^{-6}$, one cannot obtain a satisfactory solution
if $\delta_s = \delta_d$.  In that case, the value $|r|=0.068 \pm 0.034$ from
Eq.\ (\ref{eq:absr}) is too small to account for $R_d$ and $R_s$.  The value of
$|r|$ is increased to 0.100 (or larger) if we increase $R_s$ by a factor 1.4
(or larger). With such values of $r$ one may obtain the central value of $R_d$
and a suitable value of $R_s$ using Eqs.\ (\ref{eq:rd}) and (\ref{eq:rs}).
This is the essence of the need for either a larger $\b(B_s \to K^- \pi^+)$
or a larger $\xi$.  Indeed, Fleischer \cite{Fleischer:2007hj} obtained a
solution with $\delta_s - \delta_d \simeq 10^\circ$ by increasing $R_d$ and
$R_s$ by $1\sigma$.

Current data thus call for SU(3) breaking in amplitudes at the level of $20\%$
or very different strong phases.  As shown in Fig.\ \ref{fig:rACP}, both $r$
and $A_{CP}(B_s \to K^- \pi^+)$ decrease with increasing $\b(B_s \to K^-
\pi^+)$.  These conclusions are qualitatively unchanged if we allow $\delta_d$
and $\delta_s$ to differ by $\lsim 10^\circ$ for small SU(3) breaking.

\begin{figure}[t!]
\centering{\includegraphics[width=0.56\textwidth]{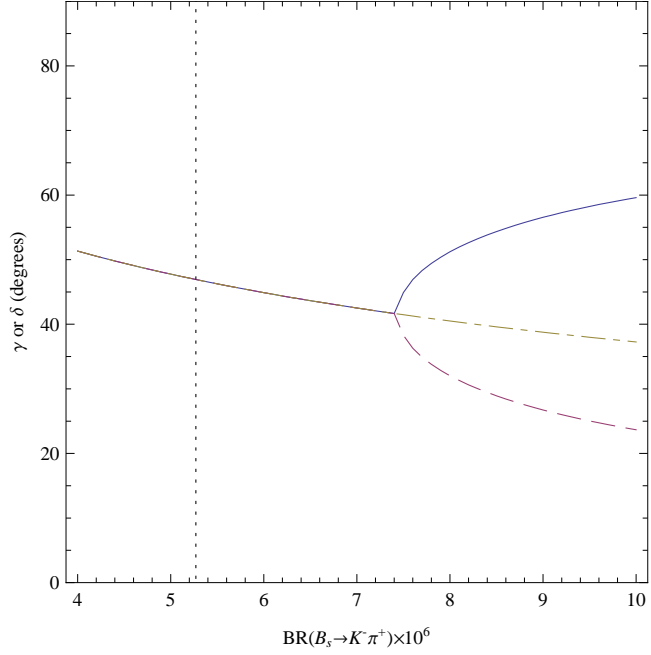}}
\caption{Solutions as function of $\b(B_s \to K^- \pi^+)$, for $r<0$ and
  $\delta_d = \delta_s \equiv \delta$.  The fork point corresponds to $\b(B_s
  \to K^- \pi^+) \simeq 7.5 \times 10^{-6}$.  The solid and dashed curves
  represent $\gamma$ and $\delta$, respectively, as preferred by other
  analyses.  A saddle point solution with $\delta_s = \delta_d$ and small
  nonzero $\chi^2$ is indicated by the dash-dotted curve.  The vertical dotted
  line indicates the current central value of $\b(B_s \to K^- \pi^+)$.}
\label{fig:gamdel}
\end{figure}

We show the dependence of $\gamma$ and $\delta$ on $\b(B_s \to K^-
\pi^+)$ in Fig.~\ref{fig:gamdel}.  Their values coincide with each
other for small values of $\b(B_s \to K^- \pi^+)$, and start to split
into three curves when it is greater than $7.5 \times 10^{-6}$.  This
occurs when $\chi^2_{\rm min}$ becomes zero for the upper
(solid) and lower (dashed) branches.  For the dash-dotted branch in
the middle, $\gamma$ and $\delta$ still coincide with each other and
continue to decrease with $\b(B_s \to K^- \pi^+)$.  The $\chi^2_{\rm
 min}$ values along this branch are small but non-vanishing, corresponding
to a ``saddle'' region in parameter space.  The upper and lower branches can
represent either $\gamma$ or $\delta$ due to the $\gamma \leftrightarrow
\delta$ symmetry.  However, the weak phase given by the solid curve is more
consistent with other analyses.  In that case, the corresponding
strong phase is given by the dashed curve.  As shown in the plot,
$\gamma$ ($\delta$) grows (decreases) monotonically with
$\b(B_s \to K^- \pi^+)$ above the fork point.

We now let $\delta_d \ne \delta_s$, permitting a test of the SU(3) symmetry
assumption.  With four observables $R_d$, $R_s$, $A_{CP}(B^0 \to K^+ \pi^-)$,
and $A_{CP}(B_s \to K^- \pi^+)$, one can solve for all four parameters $r$,
$\gamma$, $\delta_d$ and $\delta_s$ in the decay amplitudes.


\begin{figure}[t!]
\centering{
\includegraphics[width=0.46\textwidth]{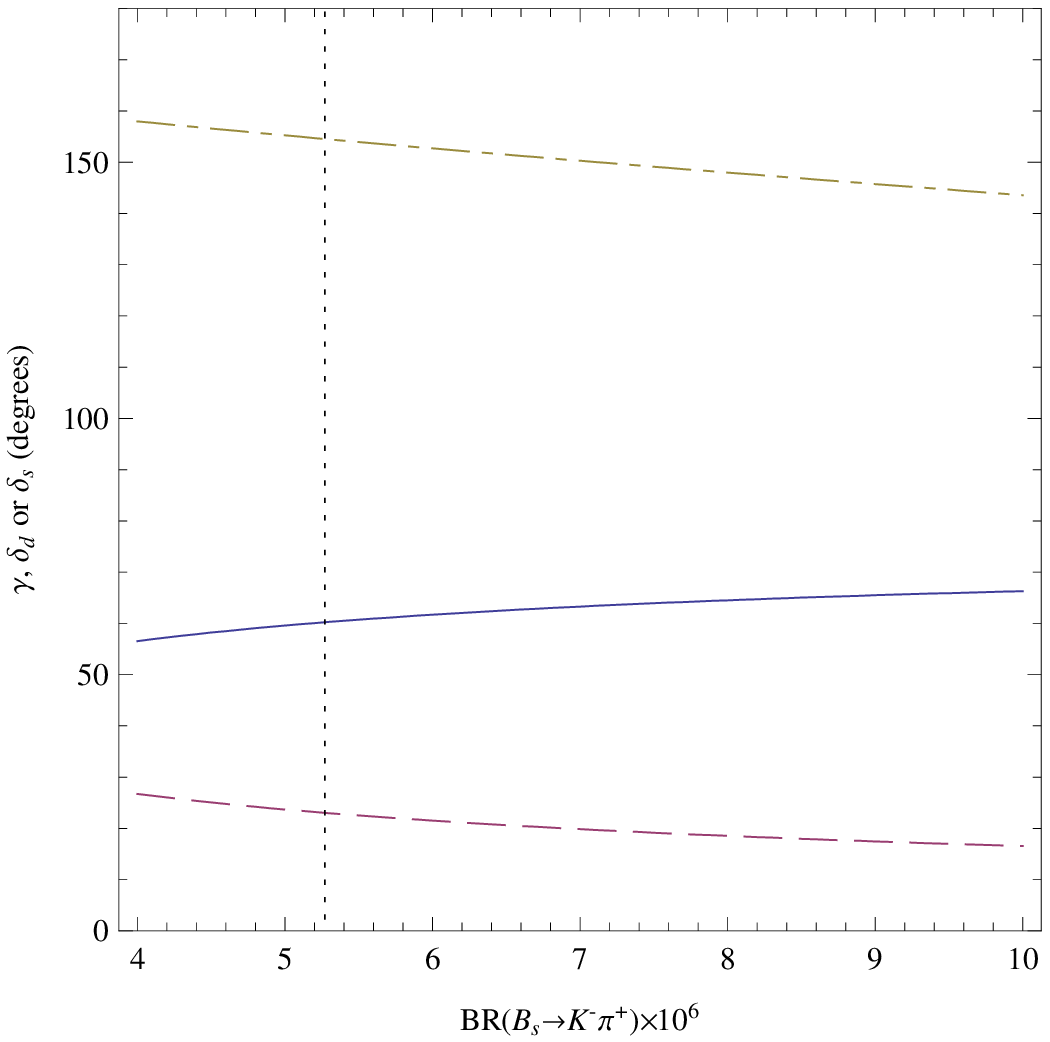} \hskip 0.2in
\includegraphics[width=0.46\textwidth]{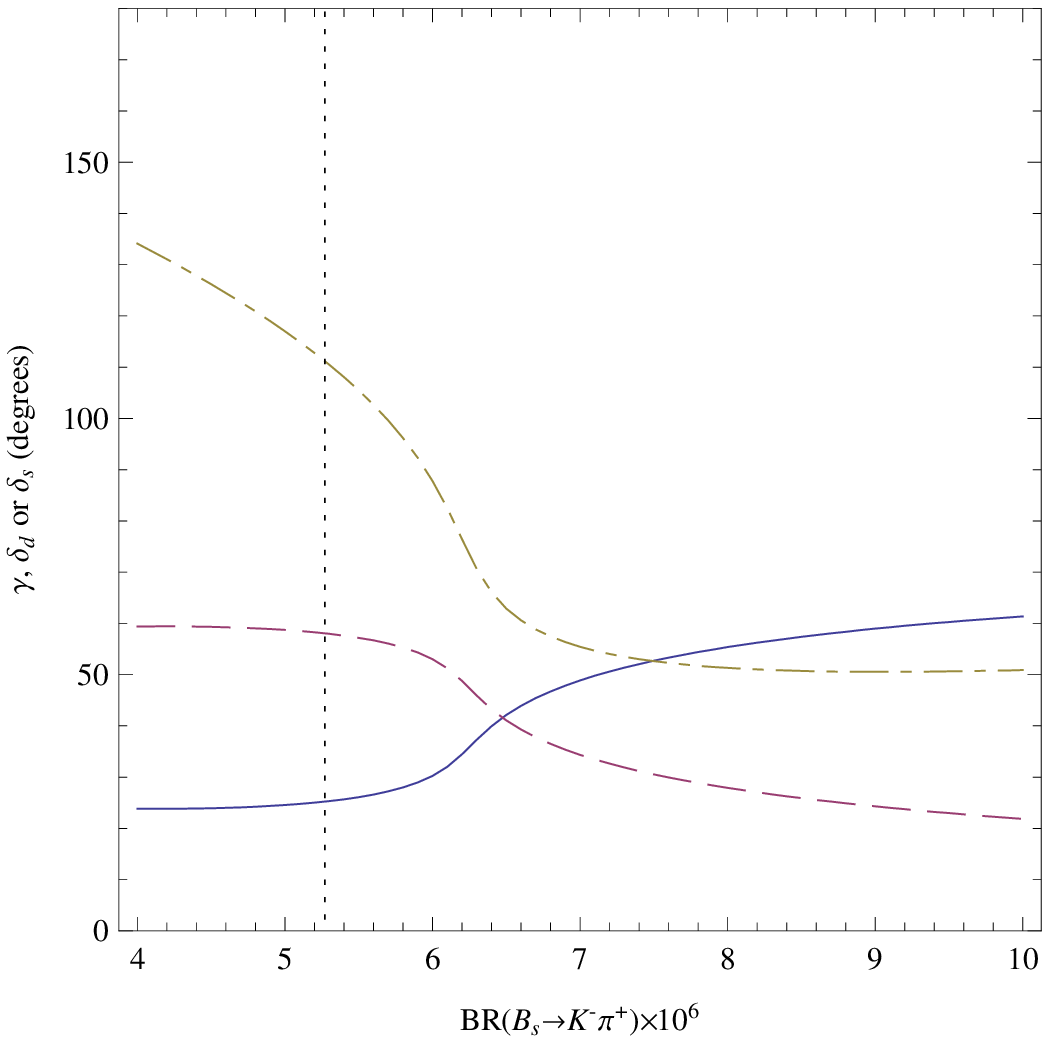}
}
\caption{Behavior of solutions as a function of $\b(B_s \to K^- \pi^+)$,
assuming $r<0$.  There are two sets of solutions (left and right)
when $\delta_d$ and $\delta_s$ are treated as independent parameters.  The
solid, dashed and dash-dotted curves represent $\gamma$, $\delta_d$ and
$\delta_s$, respectively.  The vertical dotted line indicates the
current central value of $\b(B_s \to K^- \pi^+)$.}
\label{fig:gamdel4}
\end{figure}

As shown in Fig.~\ref{fig:gamdel4}, there are two sets of possible solutions
(left and right) as a function of $\b(B_s \to K^- \pi^+)$.  For the solution on
the left, even though $\gamma$ falls within the expected range, $\delta_d$ and
$\delta_s$ differ significantly from each other.  For the solution on
the right, the strong phases are also quite different and $\gamma$ is too small
when $\b(B_s \to K^- \pi^+) \lsim 6.5 \times 10^{-6}$.  However, when $\b(B_s
\to K^- \pi^+) \ge 7.5 \times 10^{-6}$, $\gamma$ becomes reasonable, $\delta_d$
is between $20^\circ$ and $30^\circ$, and $\delta_s$ approaches $50^\circ$.  As
the current measurement of the CP asymmetry of $B_s \to K^- \pi^+$ has an error
over $40\%$, we expect it to have a weaker constraint on the parameters,
$\delta_s$.  For the current data, two solutions are found, corresponding to
the parameters:

\begin{eqnarray}
(r,\gamma,\delta_d,\delta_s) &=& (-0.128,60^\circ,23^\circ,155^\circ)
~, 
\nonumber \\
(r,\gamma,\delta_d,\delta_s) &=& (-0.121,25^\circ,58^\circ,111^\circ) ~.
\label{eq:sols}
\end{eqnarray}
In the former, $\gamma$ is more consistent with results using other methods
(for example, adding information based on $B^0 \to \pi^+ \pi^-$
\cite{Gronau:2007af}), and a small strong phase $\delta_d$ as expected in
perturbative QCD \cite{BSS,BBNS}.  However, the strong phase $\delta_s$ in both
solutions is unexpectedly large.  The $1\sigma$ ranges around the former are
\begin{eqnarray}
-0.143 \lsim r \lsim -0.112 ~,~~
& 47^\circ \lsim \gamma \lsim 72^\circ ~. &
\end{eqnarray}
The result for $|r|$ here is larger than that from Eq.~(\ref{eq:absr}) with
$\delta_d = \delta_s$.

Even though we no longer have the symmetries between the weak and
strong phases mentioned before because of the introduction of an
additional strong phase $\delta_s$, we still obtain two possible
solutions roughly corresponding to $\gamma \leftrightarrow
\delta_d$.  Within this set of observables, it is impossible to
resolve the two-fold ambiguity without resorting to some other methods
or observables.

For the solutions in Eq.~(\ref{eq:sols}), $\delta_s$ is very different from
$\delta_d$, contrary to the SU(3) symmetry assumption.  More likely
possibilities are a $B_s$ branching ratio larger than the current value or
a value of $\xi$ larger than the factorization estimate given above.
These alternatives are impossible to distinguish from one another as the
parameters $\xi$ and $R_s$ always appear in the combination $\xi^2 R_s$ [even
in $\xi^2 A_s = \xi^2 R_s A_{CP}(B_s \to K^- \pi^+)$].  A larger left-hand side
of Eq.~(\ref{eq:rs}) would entail $\cos\delta_s >0$ rather than the current
situation, permitting $\delta_s$ to be closer to $\delta_d$.  With $\xi = 1.2$,
one would obtain a solution $r = -0.11$, $\gamma = 56^\circ$, $\delta_d =
28^\circ$, and $\delta_s = 51^\circ$.  The reason that $\delta_s - \delta_d$ is
still as large as $23^\circ$ is because of the pull from $A_{CP}(B_s \to K^-
\pi^+)$.  As shown in Fig.~\ref{fig:rACP}, a smaller asymmetry is
preferred if one hopes to have $\delta_s \simeq \delta_d$.

Even though one often assumes the same SU(3) breaking factor for the tree
and penguin amplitudes, they can {\it a priori} scale differently.  Denote the
scaling factors associated with $T$ and $P$ by $\xi_T$ and $\xi_P$,
respectively.  By fixing $\xi_T = \xi$ and allowing $\xi_P$ to vary around 1,
we find that for $\xi_P \ge 1.2$ the strong phase $\delta_s$ can lie in the first
quadrant, but is still too large ($> 70^\circ$).  The weak phase $\gamma$ also
falls below $50^\circ$ in this case.  However, if we fix $\xi_P = \xi$ instead
and vary $\xi_T$, the solution improves with increasing $\xi_T$.  Taking $\xi_T
= 1.5$ as an example, we find $r = -0.129$, $\gamma = 60^\circ$, 
$\delta_d = 23^\circ$, and $\delta_s = 41^\circ$.  This shows that the scaling
behavior of $T$ plays a more dominant role.

Next, we allow both $\xi_T$ and $\xi_P$ to vary by including $\gamma =
(67.6 \pm 4.5)^\circ$ \cite{CKMfitter} obtained from other methods as
another observable constraint.  We find that if $\delta_s
- \delta_d \gsim 20^\circ$, it is possible to obtain a perfect fit to
the data.  In these cases, the preferred values of $r$, $\gamma$,
and $\delta_d$ become fixed at $-0.182$, $67.6^\circ$, and
$15^\circ$.

\begin{figure}
\centering{
\includegraphics[width=0.6\textwidth]{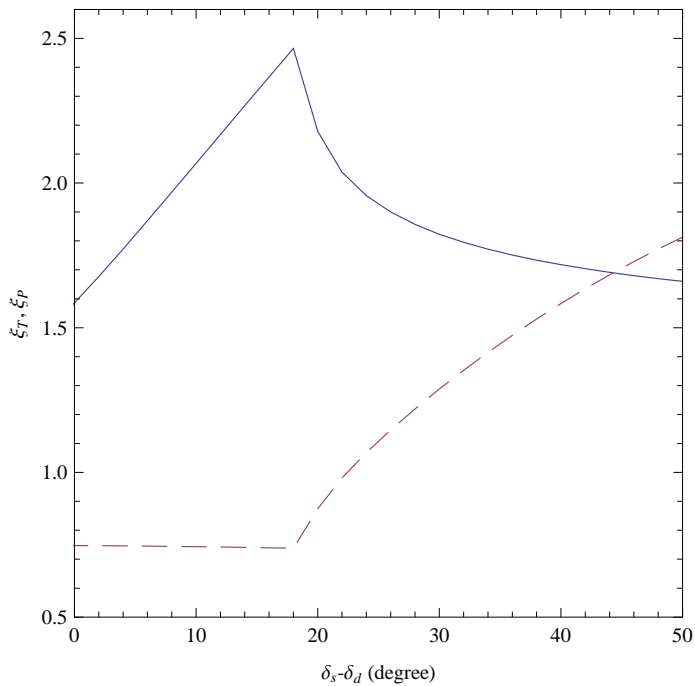}
}
\caption{Dependence of preferred values of $\xi_T$ (solid) and $\xi_P$
  (dashed) on the strong phase difference $\delta_s - \delta_d$.}
\label{fig:xiangle}
\end{figure}

The preferred values of $\xi_T$ and $\xi_P$ as a function of $\delta_s
- \delta_d$ are shown in Fig.~\ref{fig:xiangle}.  When $\delta_s -
\delta_d \lsim 20^\circ$, no perfect solution exists.  But the most
favored $\xi_T$ increases linearly with the strong phase difference,
while $\xi_P$ stays almost as a constant.  If the equality between the
two strong phases is imposed, we find $\chi^2_{\rm min} = 0.82$ with
$r = -0.114$, $\gamma = 67^\circ$, $\delta_d = \delta_s = 26^\circ$,
$\xi_T = 1.59$, and $\xi_P = 0.75$.  When $\delta_s - \delta_d \gsim
20^\circ$, $\xi_T$ drops while $\xi_P$ increases.

In Table \ref{tab:comp} we compare pairs of solutions for $\b(B_s \to K^-
\pi^+) = 7.5 \times 10^{-6}$ and $10^{-5}$ with those for the current value
$\b(B_s \to K^- \pi^+) = (5.27 \pm 1.17) \times 10^{-6}$,
keeping the same $22\%$ error.  As $\b(B_s \to K^- \pi^+)$
increases, the values of $\gamma$ and those of $\delta_d$ in
the two solutions become closer to each other.  However, the values of
$\delta_s$ remain significantly different from $\delta_d$.

To summarize, the U-spin relation between $B^0 \to K^+ \pi^-$ and $B_s
\to K^- \pi^+$ \cite{Gronau:2000md} has been utilized to obtain a
range of values of the CKM phase $\gamma$, thanks to new data on the
decay $B_s \to K^- \pi^+$ obtained by the CDF Collaboration
\cite{CDF}.  Values of $\gamma$ consistent with other determinations
and strong phases $\delta_d$ and $\delta_s$ not differing
substantially from one another may be obtained only if the branching
ratio $\b(B_s \to K^- \pi^+)$ is at least 42\% larger than its
currently quoted value of $(5.27 \pm 1.17) \times 10^{-6}$, or if the
parameter $\xi$ [Eq.\ (\ref{eqn:xi})] describing the ratio of decay
constants and form factors is more than about 1.2 (vs.\ its nominal
value of $0.97^{+0.09}_{-0.11}$).


\begin{table}
\caption{Comparison of solutions for various values of ${\b}_s \equiv \b(B_s
\to K^- \pi^+)$
\label{tab:comp}}
\begin{center}
\begin{tabular}{c r r r r r r r r} \hline \hline
${\b}_s$ & \multicolumn{4}{c}{Solution 1} & \multicolumn{4}{c}{Solution 2} \\
($10^{-6}$) & $r~~$ & $\gamma~$ & $\delta_d~$ & $\delta_s~$
            & $r~~$ & $\gamma~$ & $\delta_d~$ & $\delta_s~$ \\ \hline
 5.27 & --0.128 & $60^\circ$ & $23^\circ$ & $155^\circ$
     & --0.121 & $25^\circ$ & $58^\circ$ & $111^\circ$ \\
 7.5 & --0.148 & $64^\circ$ & $19^\circ$ & $149^\circ$
     & --0.108 & $53^\circ$ & $31^\circ$ & $53^\circ$ \\
10.0 & --0.167 & $66^\circ$ & $17^\circ$ & $144^\circ$
     & --0.133 & $61^\circ$ & $22^\circ$ & $51^\circ$ \\ \hline \hline
\end{tabular}
\end{center}
\end{table}

For the nominal values of $\b(B_s \to K^- \pi^+)$ and $\xi$, one obtains a
solution with a two-fold ambiguity, whose value of $\gamma$ in the solution
closer to other determinations (using such processes as $B^0 \to \pi^+ \pi^-$
\cite{Gronau:2007af}) is $\simeq 60^\circ$.  In this solution, however, the
strong phases are $\delta_d \simeq 23^\circ$ and $\delta_s \simeq 155^\circ$.
The latter is inconsistent with perturbative QCD calculations and its large
difference from $\delta_d$ would signal significant SU(3) breaking or failure
of factorization.  Solutions with smaller SU(3) breaking, such as those which
would result if $\b(B_s \to K^- \pi^+)$ were at least 42\% larger than its
nominal value, would be suggested if recent evaluations of $b$ quark
fragmentation \cite{HFAG,Aaltonen:2008zd} had overestimated the fraction of
$b$ quarks ending up as $B_s$.  Alternatively, the SU(3) breaking factor
$\xi$ could be larger than estimated, or could differ in tree and penguin
amplitudes.  Further studies of the $B_s \to K^- \pi^+$ decay and $b$
fragmentation at the Fermilab Tevatron and at LHCb may help to illuminate this
question.

{\it Acknowledgments:} C.-W. C. would like to thank the hospitality of the
Fermilab Theory Group during his visit where this work was started.  M. G. is
grateful to Michael Peskin and David MacFarlane for their kind hospitality 
at SLAC. We thank Michael Morello and Giovanni Punzi for helpful communications
regarding Ref.\ \cite{CDF}.  This work was supported in part by the National
Science Council of Taiwan, R.O.C. under Grant No.~NSC 96-2112-M-008-001, and by
the United States Department of Energy through Grants No.\ DE FG02 90ER40560
and DE AC02 76SF00515.

\end{document}